\documentclass[aps,prl,twocolumn,superscriptaddress]{revtex4}

\usepackage{graphicx}
\usepackage{dcolumn}
\usepackage{bm}

\begin{document}
\newcommand{\eqbeg}{\begin{equation}}
\newcommand{\eqend}{\end{equation}}
\newcommand{\vhat}{\hat{V}}
\newcommand{\ddt}{\frac{d}{dt}}
\newcommand{\dddt}{\frac{d^2}{dt^2}}
\newcommand{\toh} {t_0}
\newcommand{\tone} {t_1}
\newcommand{\tonem} {t_1^-}
\newcommand{\tonep} {t_1^+}
\newcommand{\ttwo} {t_2}
\newcommand{\tmin} {t-t_0}
\newcommand{\phis} {\varphi_s}
\newcommand{\phim} {\varphi_m}
\newcommand{\mone} {$\mbox{m}_1$}
\newcommand{\mtwo} {$\mbox{m}_2$}
\newcommand{\mrat} {m$_B$/m$_I$}
\newcommand{\mbee} {m$_B$}
\newcommand{\meye} {m$_I$}
\newcommand{\mbees} {m$_B$ }
\newcommand{\meyes} {m$_I$ }
\newcommand{\barium} {$^{136}$Ba$^+$}
\newcommand{\bariums} {$^{136}$Ba$^+$ }
\newcommand{\bars} {Ba$^+$ }
\newcommand{\levy} {L\'{e}vy }


\title{Power-law distributions in a trapped ion interacting with a classical buffer gas}


\author{Ralph G. DeVoe}
\affiliation{Physics Department, Stanford University, Stanford, California 94305}


\date{\today}

\begin{abstract}

Classical collisions with an ideal gas generate non-Maxwellian distribution functions for a single ion in a radio frequency ion trap. The distributions have power-law tails whose exponent depends on the ratio of buffer gas to ion mass.  This provides a statistical explanation for the previously observed transition from cooling to heating. Monte Carlo results approximate a Tsallis distribution over a wide range of parameters and have \it ab initio \rm agreement with experiment.
 \end{abstract}


\maketitle



The behavior of ions in the collision-free regime of a radio frequency ion trap is well understood. Laser-cooling and the properties of the quantum mechanical ground state \cite{review} have been examined in great detail. It is therefore surprizing that the more accessible regime of ions cooled by buffer gas collisions has never been thoroughly analyzed. Dehmelt first showed in 1968 that collisions with neutral buffer gases could either cool or heat the ion \cite{deh2}, depending on their relative masses. His theory, still widely accepted today, hypothesized that the recoil from a collision with a heavy neutral atom heated the ion by disrupting its response to the rf field (i.e. micromotion). The theory relied on the "pseudopotential" or time-averaging approximation and did not address the statistics of the ion's distribution function, which were assumed to be non-critical. Subsequent workers have introduced Maxwell-Boltzmann (MB) statistics in several ways, for example, by assuming a Gaussian velocity distribution function \cite{shimizu} or by hypothesizing Gaussian random noise in a Langevin equation \cite{blatt}. Numerical work has shown apparent instability in individual trajectories without addressing the statistics\cite{isolde}.

In this Letter we compute the ion's distribution function using a combination of Monte Carlo and analytic methods. Our results show that the distribution is not in general Gaussian and does not follow MB statistics. Collisions with heavy neutrals give the distribution function power-law tails  E$^{- \alpha}$ in place of the Gaussian equivalent $\exp(-E/kT)$. Here E is a time-averaged "pseudoenergy" which is not conserved during collisions\cite{deh2}. In previous work the instability was thought to arise from a positive heating rate dE/dt $>$0, leading to exponential runaway. Our results lead to a  different picture in which stationary power-law tails lead to a small but constant rate of ion loss. This leads to orders-of-magnitude differences in predicted ion lifetimes.

In the last decade non-Gaussian statistics have been studied in many different contexts \cite{tsallis}. In atomic physics such distributions have been observed primarily in laser-cooled atoms operating near the quantum-mechanical ground state. Subrecoil laser cooling has been shown to obey \levy statistics \cite{recoil_book} with a nonstationary distribution and nonergodicity \cite{recoil,ergodic}. \levy walks, anomalous diffusion\cite{walther}, and a tuneable Tsallis distribution \cite{renzoni} have been observed in optical lattices. The present case is remarkable in that it shows non-Gaussian statistics in  a classical system not far removed from the ideal gases originally studied by Maxwell and Boltzmann.

An additional motivation for understanding collisional heating is that ion traps have recently come into use as probes of collision physics. Hybrid traps comprising both ion traps and MOT's (magneto-optical trap) have been constructed yielding results for charge exchange cross-sections \cite{vuletic} and radiative lifetimes \cite{smith}. The theory of ultra-cold atom-ion collisions has been developed \cite{dalgarno,zoller_07,julienne} and novel effects of an ion in a BEC (Bose-Einstein condensate) have been predicted \cite{lukin}.  Room temperature buffer gases have been used to study molecular ions \cite{drewsen} and the properties of multipole traps \cite{wester}. This is in addition to the more traditional use of buffer gase cooling in trace element detection \cite{jesse, bjorn}, the trapping of radioactive ion beams \cite{isolde}, and several other applications. It is necessary to understand collisional heating to disentangle the effect of the trap fields from the collision physics.

Previous Monte Carlo work has used numerical integration to compute the ion's motion between collisions. This is neither fast nor accurate enough for the total of $\approx 10^{10}$ collisions needed to compute the distribution function. Instead we use the
classical time-evolution matrix M of the ion
\eqbeg
  \pmatrix{x(t_2) \cr v(t_2) \cr }= M(t_2,t_1)
    \pmatrix{ x(t_1) \cr v(t_1) \cr} 
\eqend
to propagate the position and velocity of the ion from one collision to the next.
Consider first the case of a simple harmonic oscillator, which obeys
$\ddot{x}(t) + \beta^2 x(t)=0$. The time-evolution matrix S of this system is
\eqbeg
     \text{S}(t_2,t_1)=
 \pmatrix{ \cos \beta ( t_2-t_1) & \frac{1}{\beta} \sin \beta(t_2-t_1) \cr   
                              -\beta \sin \beta (t_2-t_1) &  \cos \beta (t_2-t_1) \cr }
\eqend
. 
This is a special case of the general solution\cite{boyce}
\eqbeg
    M(t_2,t_1)= \frac{1}{D}
    \pmatrix{S_1 P_2-Q_2 R_1 & P_1 Q_2 - P_2 Q_1\cr
	    S_1 R_2 -S_2 R_1 & S_2 P_1 - Q_1 R_2 \cr}
\eqend
where P and Q are the two linearly independent solutions of a second order linear differential equation, R and S are the respective time derivatives, and the Wronskian D=S$_1$P$_1$-Q$_1$R$_1$. The rf ion trap obeys a Mathieu equation
\eqbeg
   \frac{d^2 x }{dt^2} +( a- 2 q \ \cos 2t) x = 0
\eqend 
for which  P and Q are given by the Fourier solutions\cite{review,deh2}
\eqbeg
     \mbox{P$_i$} = \sum_{m=-\infty}^{m=\infty} \cos[ (\beta +2m)t_i)] C_{2m} 
\eqend
where Q$_i$ has $\sin[ (\beta +2m)t_i)]$ in place of the cosine.
Here q= $2 e V_0 / m_i \Omega^2 r_0^2 $, a= $4 e U_0 / m_i \Omega^2 r_0^2 $, $E_0$ and $U_0$ are the rf and dc applied potentials, r$_0$ is the trap radius, and the unit of time is $2/ \Omega$, where $\Omega$ is the angular frequency of the applied rf.
The coefficients up to and including C$_{\pm 8}$ are evaluated to 24 bit accuracy ($5 \times 10^{-8}$ error) with a recursive routine. Error propagation has been tested with the identity $M(t_N,t_1)$ = $\prod_{i=1}^{N-1}  M(t_{i+1},t_i) $. For N up to $10^6$ the discrepancy $< 1 \times 10^{-5}$ for randomly chosen times t$_i$.

\begin{figure}
\includegraphics[scale=0.8]{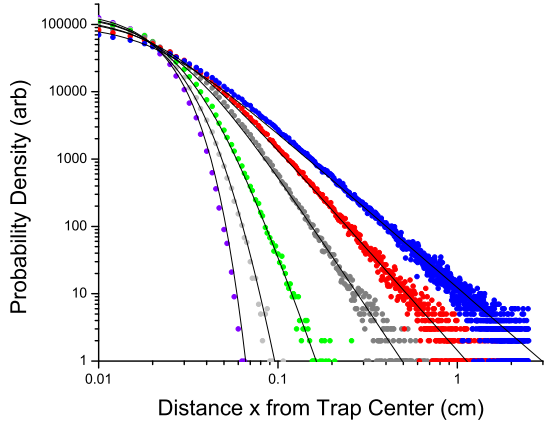}
\caption{\label{fig:epsart} Monte Carlo distributions for a single \bariums ion cooled by six different buffer gases at 300K ranging from \mbee=4 (left) to \mbee=200 (right). Note the evolution from Gaussian to power-law (straight line) as the mass increases. The solid lines are Tsallis functions Eq. 7 with fixed $\sigma$ = 0.0185 cm and the exponents of Table 1. }
\end{figure}

The distribution function is computed by Monte Carlo averaging over six random variables for each collision: the time $t_i$, the center-of-mass angles $\theta_i$ and $\varphi_i$, and the three random velocities $v^x_i$,$v^y_i$,and $v^z_i$ of the buffer gas, which obey a Maxell-Boltzmann distribution at temperature T, where T=100K, 300K, or 1000K. We assume a linear ion trap with transverse rf confinement and a DC potential along the z-axis, represented by the Mathieu matrices M$_x$ and M$_y$, where $q_y=-q_x$, and static harmonic oscillator matrix $S_z$ as in Eq. 2 above. We combine the three matrices for the x,y, and z axes into a single 6 by 6 matrix equation 
 \eqbeg
    \pmatrix{r_x(t_N) \cr r_y(t_N) \cr r_z(t_N) \cr} =  \prod_{i=1}^N C(\theta_i,\varphi_i,\vec{v_i}) \vec{M}(t_i,t_{i-1})
							\pmatrix{r_x(t_0) \cr r_y(t_0) \cr r_z(t_0) \cr } 
\eqend
where $\vec{M}$ has $M_x,M_y$ and $S_z$ along the diagonal, and where, for example, $r_x$ is the column vector $(x,\dot{x})$. Collisions are represented by a matrix C which leaves the coordinates unchanged but transforms the velocities according to a hard sphere (isotropic in the center-of-mass) collision model. The ion-neutral atom collisions are modeled by Langevin scattering in which the cross-section $\sigma \propto $ 1/v so the time between collisions is a random variable independent of the relative velocity v. An ensemble typically consists  of 10$^6$ trials each containing 500 to 50,000 collisions. All distributions for $>$500 collisions agree within statistics.

Distribution functions for a single \bariums ion at q=0.1 are shown in Fig. 1. Six different buffer gases with masses \mbees = 4, 40, 84, 136, 170 and 200 amu have been assumed, corresponding to the noble gases He, Ar, Kr, Xe as in a recent experiment \cite{jesse,bjorn} and to two hypothetical heavier gases. The distribution for \bariums in He is a good fit to an MB distribution with a classical $\sigma \approx $ $\sqrt{2kT/m_I \omega^2}$, where $\omega$ is the secular frequency $\beta \Omega$/2. All of the other gases show non-Gaussian distributions which develop broad power-law tails as the mass increases. The four heaviest gases fit a power-law $x^{-2 \alpha}$ with good $\chi^2$ over at least 3 orders of magnitude. For \mbees = 84, 136, 170, and 200 the best fits are $\alpha$ =3.2, 1.98, 1.5, and 1.17 respectively. A typical fit will have $\chi^2/n \le 1.1 $ for n$> 100$ degrees of freedom.

\begin{table}
\caption{Tsallis parameters n and $q_T$ fit from Fig. 1}
\vspace{.1in}
\begin{ruledtabular}
\begin{tabular}{cccc}
  Buffer Gas  & \meye/\mbee  & n  & q$_T$ \\ \hline

 He & 34.5    &  $>$60 &  1.03 \\
 Ar & 3.40   &  8.2 & 1.12 \\
 Kr &  1.70 & 3.8 &  1.26  \\
 Xe &  1.0  & 1.98 &  1.51 \\
 170  &   0.80    & 1.50 & 1.80 \\
 200 &   0.68     &  1.15 & 1.87
\end{tabular}
\end{ruledtabular}
\end{table}

In the absence of an analytic theory, the data has been fit to a Tsallis function $T(x/\sigma,n)$ \cite{tsallis}
\eqbeg
          T(x/\sigma,n) = \frac{T_0}{[1+(x/\sigma)^2/n]^n}
\eqend
which is a generalization of the Gaussian. For  n $\rightarrow \infty$ T reduces to a Gaussian while for small n it has power-law tails of the form $(x/\sigma)^{-2n}$. The exponent n is related to the more familiar "entropic" Tsallis parameter q$_T$ by q$_T$=1+1/n . The Tsallis function arises in the theory of nonextensive entropy\cite{tsallis} but at present we treat it empirically. Table 1 shows the value of n extracted from fitting the distributions to Eq. 7, where $\sigma$ was held constant for all \mbees and T$_0$ normalizes the distribution to unity. The fit is qualitative since the $\chi^2$ is poor due to systematic deviation near the origin, where the standard deviation $< 0.3 \%$. Nevertheless Fig. 1 shows good agreement over a factor of $10^5$ in probablility density and a factor of 50 in buffer gas mass. The value of the Tsallis exponent n is close to the value of $\alpha$ extracted from the power law fit above. Similar data at 100K and 1000K give comparable fits with the same $\sigma$ scaled by $\sqrt{T}$. 

The usefullness of the Tsallis function is that it shows that n and $\sigma$ act independently of each other, to first order. In the light gas limit \mbees $\rightarrow 0$, the ion has a Gaussian distribution with a temperature T equal to that of the buffer gas, so that $\sigma \approx $ $\sqrt{2kT/m_I \omega^2}$. As \mbees increases, $\sigma$ changes very slowly, so that the distribution retains a Gaussian-like core of constant width as the power-law tails get stronger. This indicates that the increase in the mean energy of the ion is not the cause of ion loss. A three-parameter fit, in which $\sigma$, n, and the normalization are optimized for each value of \mbee, shows a weak dependence of $\sigma$ on \meye/\mbee. For example, the best value of $\sigma$=0.0175 cm at \mbee=4  rises to $\sigma$=0.022 cm at \mbee=200, an increase of only 26 $\%$ for a 50-fold decrease of \meye/\mbee. Similarly, changing the temperature of the buffer gas does not alter the power-law exponent. To generalize further, the Tsallis exponent n  is approximated by the simple relation n $\approx$ 2\meye /\mbees, which is accurate in the exponent to about $\pm 20 \%$. 

The Monte Carlo also computes the ion lifetime. The ion is started at the origin with zero energy and is propagated through i collisions, until r$_i$= $\sqrt{x_i^2+y_i^2}$ $\ge$ r$_0$, the trap radius. In general the ion lifetime $\tau \propto r_0^{2n}$, where n is the Tsallis exponent of Table 1. However, since the trap depth U $\propto$ r$_0^2$ (for constant q), it is more general to plot $\tau$ versus U, which yields $\tau \propto$ U$^n$ as shown in Fig. 2. Interestingly, $\tau$ is not sensitive to initial conditions and an ion starting with an energy $\approx$ 1 eV has $\tau$ only slightly shorter than with zero energy. This is because most of the hot ions equilibrate to 300K in a few dozen collisions. It is only in a very large ensemble ($10^6$ trials in Fig. l ) or a very large number of collisions (N=5 $\times 10^5$ in Fig. 2) that extreme values of r$_i$ are reached. In contrast to the exponential runaway model, Fig. 2 suggests that ion traps may be designed to achieve a specific ion lifetime.

\begin{figure}
\includegraphics[scale=.8]{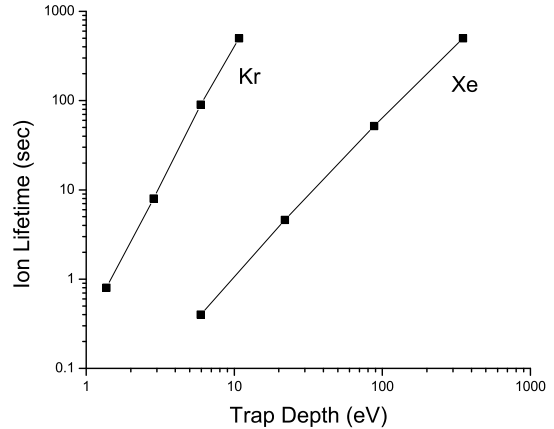}
\caption{\label{fig:epsart} Predicted power-law ion lifetime $\tau \propto$ U$^n$ where U is the trap depth, n is the Tsallis exponent of Table 1, and the parameters correspond to Fig. 1}
\end{figure}

Power-law tails dominate the trap stability whenever a stationary distribution function exists. However, when the Tsallis exponent falls below 1, which occurs for  $m_B > 1.55$ $m_I$, the distribution becomes time-dependent, as in a \levy distribution\cite{recoil_book}, the mean values diverge, and the ion's energy increases with each collision. In this case exponential runaway occurs as originally suggested in \cite{deh2}.

The Monte Carlo results have no free parameters and the predicted ion lifetimes agree with the results of a recent experiment \cite{jesse,bjorn} in which a single Ba$^+$ was confined in a trap of radius R$_T$=0.26 cm and q=0.52. Stable trapping was observed for He gas, Ar gas was measured to give an ion lifetime of 50-100 sec, while  Kr and Xe had lifetimes too short to measure ($<$ 5 sec). When the above R$_T$ and q are input to the Monte Carlo it yields lifetimes of 45 sec for Ar and $<0.1$ sec for Kr and Xe. 

It remains to provide a physical explanation of these results. An analytic expansion of the Mathieu matrix Eq. 3 and Eq. 6 shows that the power-law tails are a result of a multiplicative random process\cite{redner}, i.e., products of random variables. These may be contrasted with the better-known additive random processes (sums of random variables), which obey the central limit theorem and produce Gaussian statistics. Multiplicative random products are not well understood but they do not in general lead to Gaussian distributions. Multiplicative fluctuations have recently been studied in a Langevin equation \cite{multi} and have been shown to lead to a tuneable Tsallis distribution.

A model for multiplicative fluctuations can be derived by expanding the Mathieu matrix Eq. 3 to first order in q and substituting the result in Eq. 6. In the same limit used by Dehmelt\cite{deh2} one can show\cite{derive} that the matrix product Eq. 6 can be approximated by a product of numbers
$\prod_{i=1}^N R(\phis^i,\phim^i)$
where
\begin{widetext}
\eqbeg
      R(\phis,\phim) = 
 \sqrt{ \cos^2 \varphi_s + \alpha^2 \sin^2 \varphi_s + 2 (\alpha-1)^2 \cos^2 \varphi_s \sin^2 \varphi_m - \sqrt{2} \alpha (\alpha -1 ) \sin 2 \varphi_s \sin \varphi_m } 
\eqend
\end{widetext}
Here $\phis^i$ and $\phim^i$ are the phases of the secular motion ("macromotion") and driven rf oscillations ("micromotion") at the time of the i-th collision and $\alpha = (m_I-m_B)/(m_I+m_B)$ is a recoil parameter. Multiplicative random products tend to be dominated by rare events of large amplitude\cite{redner}. In the present case these events can be identified as N consecutive heating collisions without an intervening cooling collision. Consider a volume of phase space $\beta <1$ around the heating maximum R$\approx \sqrt{3}$ in Fig. 3. N consecutive collisions will give an amplitude of 3$^{N/2}$ with a probability of $\beta^N$. This provides a mechanism for the power-law tails leading to ion loss. If the trap were of infinite size, cooling collisions, which occupy most of the phase space, would eventually return the ion to the origin. In this sense, the ion loss is due to a non-Gaussian \it fluctuation \rm rather than to heating. In the light mass limit $\alpha \rightarrow 1$ Gaussian statistics return since each term in the product is near unity,  R= $ 1 + \epsilon(\varphi_s,\varphi_m) $ where $\epsilon << 1$. The product reduces to a sum $1+ \sum \epsilon(\varphi_s,\varphi_m)$, the fluctuations become additive, and the central limit theorem applies.

\begin{figure}
\includegraphics[scale=.8]{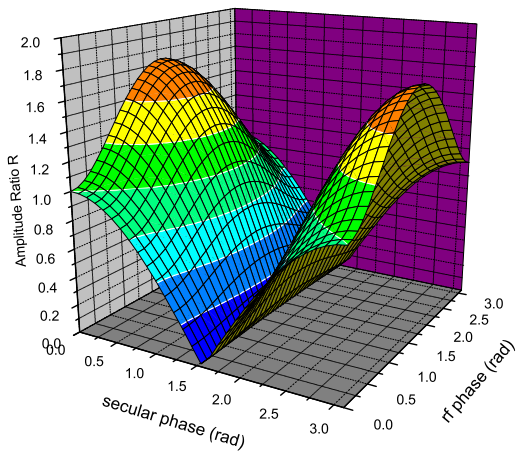}
\caption{\label{fig:epsart} A heating and cooling diagram for a single collsion in the case $m_b=m_i$ or $\alpha=0$ in Eq. 8. Heating occurs for R$>1$ and cooling for R$<1$. The explanation given in the text does not depend on the exact form of this result and requires only that there be a small phase space volume $\beta <1$ for R $>$ 1. }
\end{figure}

This work has several implications. For statistical mechanics it provides a simple, classical system which shows tuneable non-Gaussian statistics. For trapping and cooling experiments it shows how traps may be engineered for a specific ion lifetime, as in Fig. 2. This should be useful in trace atom detection\cite{jesse,bjorn} and in trapping radioactive ion beams\cite{isolde}. It is also necessary for understanding recent ion trap collision experiments ref[11-17], since the non-Gaussian distribution function can alter their interpretation.

\end{document}